\documentclass[12pt]{article}
\usepackage{graphicx}

\newcommand\pubdate{\today}

\newcommand\pubnumber{}

\def\Title#1{\begin{center} {\Large #1 } \end{center}}
\def\Author#1{\begin{center}{ \sc #1} \end{center}}
\def\Address#1{\begin{center}{ \it #1} \end{center}}

\newcommand\pubblock{\rightline{\begin{tabular}{l} \pubnumber\\
         \pubdate  \end{tabular}}}
\newenvironment{Abstract}{\begin{center}{\bf Abstract}\end{center} \bigskip \begin{quotation}  }{\end{quotation}}
\newenvironment{Presented}{\begin{quotation} \begin{center} 
             PRESENTED AT\end{center}\bigskip 
      \begin{center}\begin{large}}{\end{large}\end{center} \end{quotation}}

\def\beq{\begin{equation}}
\def\eeq#1{\label{#1}\end{equation}}
\def\eeqn{\end{equation}}

\def\beqa{\begin{eqnarray}}
\def\eeqa#1{\label{#1}\end{eqnarray}}
\def\eeqan{\end{eqnarray}}

\let\bar=\overbar

\def\Dslash{\not{\hbox{\kern-4pt $D$}}}
\def\dslash{\not{\hbox{\kern-2pt $\del$}}}

\def\msb{{\bar{\ssstyle M \kern -1pt S}}}

\begin{document}
\begin{titlepage}
\pubblock

\vfill


\Title{The LHCb Upgrade}
\vfill
\Author{P. Collins \\ on behalf of the LHCb collaboration}  
\Address{CERN, CH1211 Geneve 23, Switzerland}
\vfill


\begin{Abstract}
The LHCb detector at the LHC has shown a very successful initial operation and it is expected that the experiment will accumulate an integrated luminosity in proton-proton collisions of around $1~{\rm fb^{-1}}$ in 2011.  The data already collected are being used to pursue the experiment's primary physics goal that is the search for New Physics via the measurement of CP asymmetries and rare decays in the b and c sector.  The LHC is already capable of delivering higher luminosity than is currently used at LHCb, and an LHCb upgrade is planned for 2018 which will allow the detector to exploit higher luminosity running, and at the same time to enhance the trigger efficiencies, particularly in the hadronic decay modes.  This upgrade will allow the experiment to accumulate an integrated luminosity of $\sim 50~{\rm fb^{-1}}$ over the following decade, and acquire enormous samples of b and c hadron decays to allow for more precise measurements and a deeper exploration of the flavour sector.  In addition, the flexibility of the new proposed trigger together with the unique angular coverage of the LHCb experiment opens up possibilities for interesting discoveries beyond the flavour sector, and will allow LHCb to focus on the physics channels which will be of most interest in the light of the discoveries of the coming decade.
\end{Abstract}

\vfill

\begin{Presented}
The Ninth International Conference on\\
Flavor Physics and CP Violation\\
(FPCP 2011)\\
Maale Hachamisha, Israel,  May 23--27, 2011
\end{Presented}
\vfill

\end{titlepage}
\def\thefootnote{\fnsymbol{footnote}}
\setcounter{footnote}{0}
%


\section{Introduction}

The upgrade of the LHCb detctor is planned to take place around 2018.
The strategy for this upgrade builds on the very successful startup of the current
experiment, during which the hardware has shown an excellent
performance, the trigger strategy has been validated, and the quality of the 
physics results using first data is in line with or has exceeded the
goals of the LHCb physics roadmap documents~\cite{ref:roadmap}.
This paper is organised as follows.  Firstly the LHCb experiment, together with
its current limitations, is outlined, and the upgrade of the
trigger to a 40 MHz readout is introduced.  Then the physics programme of
the upgraded detector, which is expected to collect an integrated
luminosity of $50~\rm{fb^{-1}}$ with improved trigger efficiency,
is described.  Finally the detector environment of the upgrade and the
principal hardware interventions which are required are discussed.  
For a full description of the LHCb upgrade programme, the reader is
referred to the Letter of Intent~\cite{ref:LoI}, which has recently been approved by the LHCC.

\section{The LHCb Experiment; Performance and Limitations}

\begin{figure}[htb]
\centering
\includegraphics[width=0.7\textwidth]{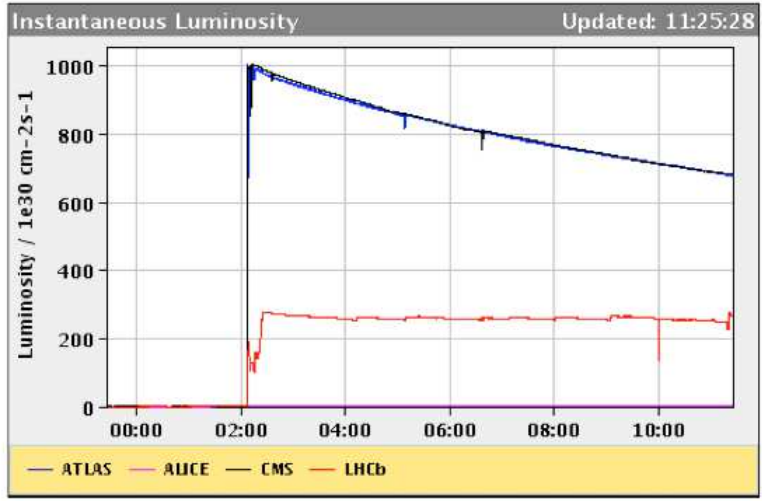}
\caption{Typical evolution of the luminosity during an LHC fill in Spring 2011, with the delivered luminosity to the general purpose detectors ATLAS and CMS shown in blue and black, and the delivered luminosity to LHCb in red.  The observed near constant luminosity at LHCb is obtained through ``luminosity levelling''. }
\label{fig:lumilevelling}
\end{figure}

LHCb is a forward spectrometer experiment measuring the products of proton-proton collisions at the LHC.  It is designed to exploit the enormous $\bar{b}b$ and $\bar{c}c$ cross section at the LHC by triggering on and reconstructing the beauty and charm meson decay products.  Its main physics goal is the indirect search for New Physics via the measurement of CP asymmetries and the search for and study of rare decay processes.   The forward geometry allows the experiment to capture 40\% of the production cross-section, and to exploit the long flight distance of $B$ hadrons for performing time dependent CP violation measurements, in particular in the $B_s$ sector.  Comprehensive particle identification, essential for disentangling the $B$ hadron decay modes, is provided by the RICH system, muon chambers and calorimeters.  LHCb was designed to run at a luminosity of $2 \times 10^{32} \rm{cm^{-2}s^{-1}}$, which is below the luminosity at which ATLAS and CMS operate.  Following trigger optimisation and an excellent detector commissioning phase the experiment is currently running  at $3 \times 10^{32} \rm{cm^{-2}s^{-1}}$ and tests at higher luminosity are envisaged.  
The differential in luminosity between LHCb and ATLAS/CMS is achieved by imposing an offset in the beams at LHCb.  This offset is reduced as the beam intensity decays, so that a constant luminosity is delivered to LHCb throughout the fill.  The results of this called luminosity levelling technique~\cite{ref:levelling} are illustrated in Fig.~\ref{fig:lumilevelling} for a typical fill.

During the LHC start-up phase in 2010 LHCb has shown an excellent performance.  Key performance indicators include a primary vertex resolution (for 25 tracks) of $\sim 13~{\rm \mu m}$ in $x$ and $y$, an impact parameter resolution of $13~{\rm \mu m} + 25/{\rm p_T}~{\rm \mu m}$, where ${\rm p_T}$ is the transverse momentum in GeV/c, and a proper time resolution of $\sim 50~{\rm fs}$.  The particle identification system is functioning well, and typical B hadron mass resolutions of 7-11 MeV have been demonstrated.   With the $\sim 36 {\rm pb^{-1}}$ recorded in 2010 the LHCb physics programme is well underway; highlights include first observations of decay modes such as $B_s \rightarrow J/\psi f_0$~\cite{ref:bsjpsif0},
$B_s \rightarrow D^0 K^{*0}$~\cite{ref:bsd0kstar}, 
$B_s \rightarrow K^{o*} \overline{K^{o*}}$~\cite{ref:bskstarkstar,ref:k0stark0starthisconference},
as reported in this conference, a non systematic limited measurement of $\Delta m_s$ at the $0.7\%$ level~\cite{ref:betasthisconference} and a limit on ${Br} (B_s \rightarrow \mu^{+} \mu^{-})$~\cite{ref:bsmumu} already reaching Tevatron sensitivity.  By summer 2011 LHCb expects to collect up to an order of magnitude greater statistics and make significant advances in the core programme of flavour physics measurements.

With the current running conditions and machine schedule it is anticipated that LHCb will collect of the order of $5~{\rm fb^{-1}}$ at centre of mass energies of 7 TeV and 14 TeV by 2016/7.  After this amount of integrated luminosity the time to double statistics becomes too slow and thus LHCb plans to upgrade the detector to enable statistics to be accumulated at a significantly enhanced rate.  This cannot be achieved by simply running at higher luminosity due to the fact that the current trigger scheme incorporates a  hardware trigger at the earliest level, which uses information from objects of high transverse energy in the muon chambers and calorimeters, and reads out information from the entire detector at a maximum rate of 1 MHz.  The consequence of this is that an increase in luminosity would force a corresponding raising of the transverses energy threshold for triggering in the hadronic channels, where the rates are very high, and there would be no net overall gain.  The time available to this first level trigger is too short to be able to implement a more complex decision, even if extra information could be made available.  It is therefore proposed to adopt a strategy of reading out the entire detector at a frequency of 40 MHz, and executing all the trigger algorithms in software, with an output rate of $\sim 20$kHz.    The event farm where this trigger would run would have to be sufficiently large to cope with the data rate, and a staged approach is anticipated, whereby a low level trigger with similar functionality to the current first level trigger, but a tunable rate, will be maintained.  This low level trigger will protect against occupancy fluctuations and also provide a throttle which enriches the events during the staging period of the computing farm.  The high level trigger algorithms will have all the event information available, and it is anticipated that a factor of up to 2 in efficiency can be achieved for the hadronic modes, which can be combined with the gain of running at increased luminosity.  In the first phase of the upgrade the target running luminosity is $10 \times 10^{32} \rm{cm^{-2}s^{-1}}$ and it is planned to accumulate $50~{\rm fb^{-1}}$ over a period of 10 years.   The evolution of signal efficiency at high luminosity running as a function of the bandwidth allocated to the LLT is shown in table~\ref{tab:results}.  It can be seen that very good efficiencies are achieved as soon as the size of the computing farm allows for high LLT rates.
In order to allow for the readout of LHCb at 40 MHz all the FE electronics must be adapted or redesigned, which also implies the complete replacement of components where the frontend (FE) electronics is integrated, such as the silicon detectors and RICH HPDs.  At the same time LHCb will take the opportunity to implement detector enhancements to cope with the increased occupancies and integrated radiation levels associated with high luminosity running.

\begin{table}[!hbtp]
\begin{center}
\begin{tabular}{l|ccc}  
\hline\hline
LLT-rate (MHz)   &  1 &  5 & 10 \\ \hline
$B_s \rightarrow \phi \phi $    &   0.12  &     0.51 & 0.82   \\
$B^0 \rightarrow K^* \mu \mu $    &   0.36  &     0.89 & 0.97   \\
$B_s \rightarrow \phi \gamma $    &   0.39  &     0.92 & 1.00   \\
\hline\hline
\end{tabular}
\caption{Signal efficiency, calculated for running at a luminosity of $10^{33} {\rm cm^{-2} s^{-1}}$, for hadronic modes of particular interest at the upgrade, as a function of the rate allocated to the tuneable Low Level Trigger.  This rate will be set to a maximum as soon as the staging of the event filter farm allows.  From~\cite{ref:LoI}.}
\label{tab:results}
\end{center}
\end{table}

\section{The LHCb Upgrade Physics case}

\begin{figure}[htb]
\centering
\includegraphics[width=0.7\textwidth]{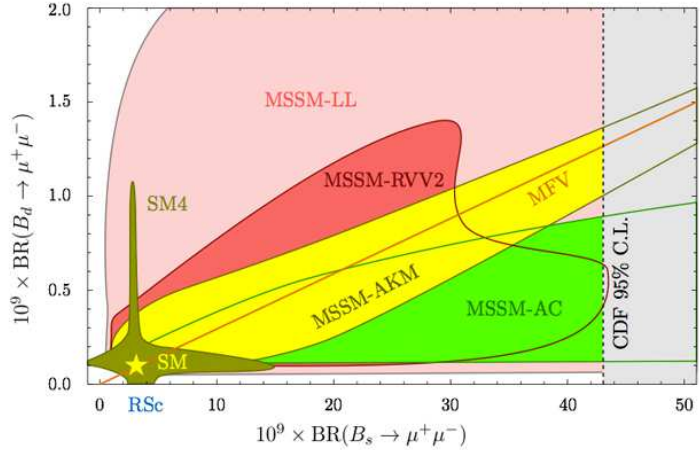}
\caption{Correlations between the branching ratios of $B_s \rightarrow \mu^+ \mu^-$ and $B_d \rightarrow \mu^+ \mu^-$ in MFV, the SM4 and four SUSY flavour models.  The gray area delineates the region excluded out by experiment and the Standard Model region is indicated by the yellow star.   Taken from~\cite{ref:straub}.}
\label{fig:mfvmssm}
\end{figure}

The primary mission of LHCb is to detect New Physics via pioneering exploratory measurements.  Should New Physics be discovered at LHCb (or elsewhere), the measurements at the upgraded experiment will be needed to distinguish between different New Physics models via correlated and more precise exploration.  If the LHCb measurements point towards the Standard Model expectations, the precision of the upgrade will be needed to uncover the correspondingly small levels where New Physics may contribute.  Two examples are given here:
\begin{itemize}
\item{The search for $B_s \rightarrow \mu^+ \mu^-$, where LHCb may observe an enhancement to the Standard Model branching ratio.  Two models  which could be at the source of such an enhancement are Minimal Flavour Violation (MFV) or a 4-family Standard Model with two new quarks and 5 new quark-mixing parameters.  In both models the branching ratio can be enhanced, but in MFV the ratio ${Br}(B_d \rightarrow \mu^+ \mu^-)/{Br}(B_s \rightarrow \mu^+ \mu^-)$, is constrained, as illustrated in figure~\ref{fig:mfvmssm}.  The task of the upgraded experiment will be to disprove or constrain such models with improved flavour data.  Observation of $B^0 \rightarrow \mu^+ \mu^-$ requires huge statistics and excellent control of backgrounds, for which the statistics accumulated by LHCb before the upgrade are unlikely to be sufficient.}
\item{Analysis of angular asymmetries in the channel $B \rightarrow K^* \mu^+ \mu^-$.  With the statistics available to LHCb the zero crossing point of the $A_{FB}$ symmetry can be measured with a precision comparable to the theoretical expectation~\cite{ref:roadmap}.  However the full power of the upgrade is needed to deploy new observables and access the full kinematic distributions, where stronger theoretical constraints exist~\cite{ref:crossingfuture}.  An example is the transversity asymmetry $A_T^{(2)}$, which is highly sensitive to new right-handed currents to which the forward-backward asymmetry is blind, and for which the statistical power of the upgrade will be needed.  The upgrade also has unique potential to explore related decays, for instance $B_s \rightarrow \phi \mu^+ \mu^-$, which can be used to measure CP violation in interference between mixing and decay amplitudes.}
\end{itemize}

In addition to New Physics searches the current experiment is designed to make precision measurements in new territory, such as a measurement of the CKM angle $\gamma$ with a precision of $\sim 3^\circ$.  Here the task of the upgrade will be to reach a sub $1^\circ$ precision, via the increased statistics and the deployment of new channels, and this will be necessary to match the anticipated improvements in theory as more powerful lattice calculations become available~\cite{ref:shigemitsu}.  

Finally, the LHCb upgrade will open up a new world of exploration in flavour physics and beyond, driven by the results which the coming years of LHC operation are expected to unveil.  The implementation of a flexible software trigger in an experiment with a very complementary geometry to ATLAS and CMS will allow the experiment to focus efficiently on the topologies of interest.

\begin{figure}[htb]
\centering
\includegraphics[width=1.0\textwidth]{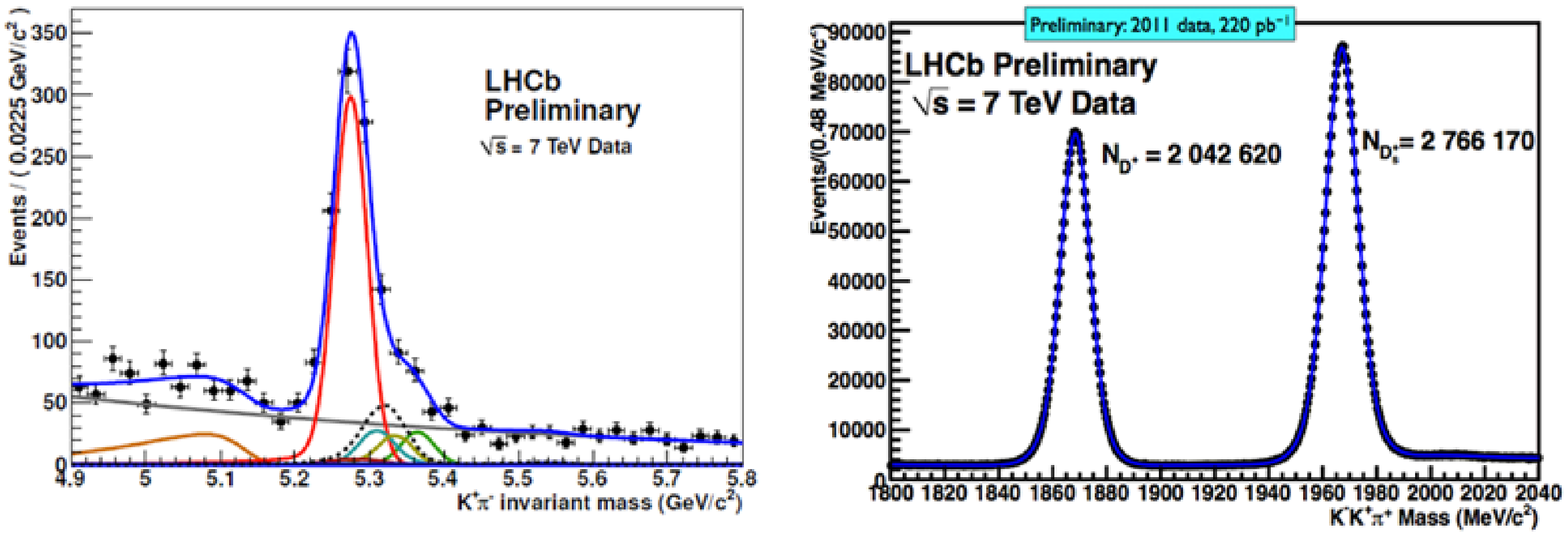}
\caption{Plots illustrating the performance of the current LHCb detector.  The left plot shows invariant mass distributions for $B^0 \rightarrow K^+ \pi^-$ candidates, illustrating the low background, excellent mass resolution and use of particle identification of the detector.  The right plot shows reconstructed $D^+$ and $D_s^+$ candidates in the $KK\pi$ modes reconstructed with $220~{\rm pb^1}$ of data, illustrating the very large statistics which are already accessible.}
\label{fig:lhcbpeaks}
\end{figure}

In this section we discuss some channels, in addition to the ones mentioned above, which demonstrate the extra physics reach brought by the upgrade.  For a more complete discussion we refer the reader to the Letter of Intent~\cite{ref:LoI}.  The capacity of the current LHCb detector to perform the measurements is illustrated in figure~\ref{fig:lhcbpeaks}

\begin{itemize}

\item{$\beta_s$ with $>> 5~{\rm fb^{-1}}$}
$B_s \rightarrow J/\psi \phi$, which gives access to a measurement of CP violation in the mixing phase, is an example of a channel with a very clean reconstruction in LHCb.  
The SM expectation is small and precisely predicted, and sensitive to NP effects.   Results presented at this conference~\cite{ref:betasthisconference,ref:betasthisconference2} have shown that an intriguing hint of a deviation from the Standard Model is seen by both the Tevatron experiments and also by LHCb, with the same sign and order of magnitude in significance.  More light may be shed on this anomaly by summer of this year, and the LHCb performance indicates that an uncertainty comparable to the SM value can be reached within $5~\rm{fb^{-1}}$, giving the possibility to either find or rule out a large effect.    The LHCb upgrade will improve this measurement with an order of magnitude increase in statistics and will also enhance it with the addition of new channels.  Pure CP states such as $B_s \rightarrow D_s^+ D_s^-$ and $B_s \rightarrow J/\psi f_o$ give alternative measurements which do not need the complicated angular analysis of the $J/\psi \phi$ two vector final state.  It is also expected that SM uncertainties due to (suppressed) penguin contributions can be attacked using $B_s \rightarrow J/\psi K^*$ and penguin-free $B_s \rightarrow D^0 \phi$ decays.  

\item{CP violation in mixing via $A_{fs}(B_s)$}

Another hint of New Physics lurking around the corner has been provided at this conference with the inclusive flavour specific asymmetry measurement from the D0 experiment which has presented a 3.2 sigma deviation from the SM of the combined $B_0$ and $B_s$ lepton asymmetry~\cite{ref:afbthisconference}.  LHCb plans to confirm or rule out this anomaly via  the difference of asymmetries of $B_s \rightarrow D_s^- \mu^+ X$ and $B_d \rightarrow D^- \mu^+ X$ with decays to the identical final state, in order to reduce biases due to detector asymmetry.    At the upgrade, due to the flexible and efficient software trigger it will be possible to access this quantity via hadronic decay channels, such as the flavour specific decay $B_s \rightarrow D_s^- \pi^+$, with $D_s^- \rightarrow K^+ K^- \pi^-$.  In this case the detector asymmetry can be suppressed due to the fully charge symmetric final state, and the statistics available at the upgrade can by fully exploited.

\item{Charmless hadronic B decays}

Charmless B decays, which proceed via rare loop penguin diagrams are very sensitive to New Physics effects, and LHCb will make the first measurements of direct CP violation in $B_s$ and $\Lambda_b$ decays and of time dependent CP violation in the decay $B_s\rightarrow K^+ K^-$ as well as the observation of new channels such as $B_s \rightarrow K^{*o} \overline{K^{*o}}$ as presented at this conference~\cite{ref:k0stark0starthisconference}.  At the upgrade it will be possible to make precision time dependent CP violation measurements in penguin dominated diagrams such as $B_s \rightarrow \phi \phi$.
This channel benefits particularly from the improved software trigger at the upgrade.  At the upgrade additional channels can be added and unique measurements can be made of multibody $B_s$ decays.

\item{Charm Physics}

With the $37 {\rm pb^{-1}}$ collected in 2010 LHCb has already accumulated charm samples of $D^o \rightarrow h^+ h^-$ decays comparable to those of the B factories.  The efficiency with the current experiment is good for 2-body decays, however the efficiency drops for multibody decays due to the lower transverse momentum of the decay products and the first level trigger limitation of the current experiment.  At the upgrade the fully software based trigger will allow selection of the topology of interest and very large data samples will be collected.  This will give sufficient statistics to allow precise studies of CP violation in mixing and decay of charmed hadrons, in addition to searches for very rare charm decays such as $D \rightarrow \mu^+ \mu^-$ and lepton flavour violation e.g. $D \rightarrow e \mu$.  

\item{Physics beyond flavour}

The forward acceptance of LHCb means that LHCb is complementary to ATLAS and CMS in many important topics beyond flavour, an attribute which is enhanced by LHCb's unique vertexing and PID capabilities.  Further advantages are expected to accrue from the fully flexible trigger and the ability to run at high luminosity.  LHCb is expected to contribute in the areas of electroweak physics, search physics and exotics, and QCD.  These topics are discussed in more detail in the Letter of Intent~\cite{ref:LoI}.

\end{itemize}

\section{Detector Upgrade Environment}

\begin{figure}[htb]
\centering
\includegraphics[width=1.0\textwidth]{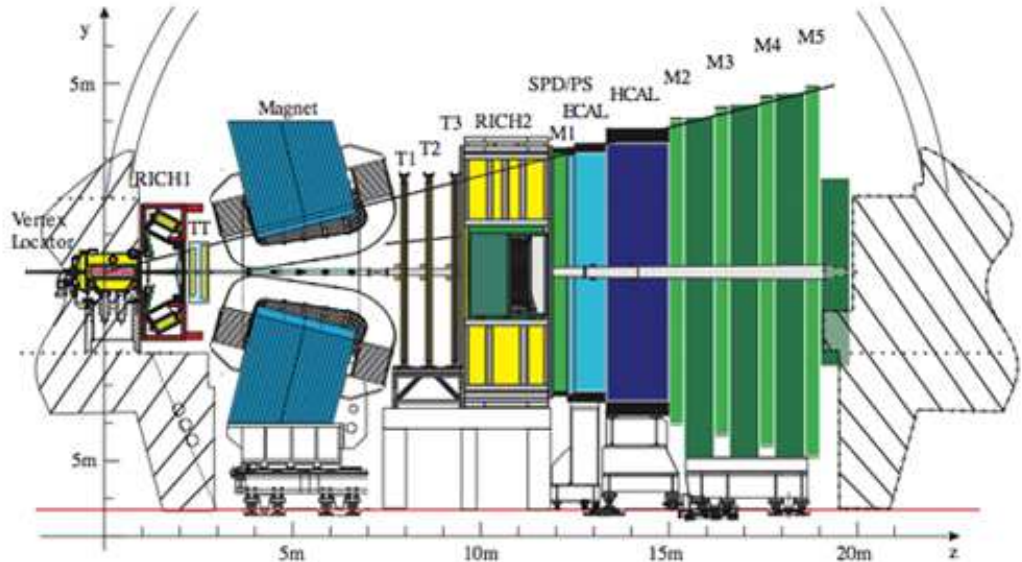}
\caption{Layout of the current LHCb detector}
\label{fig:lhcb}
\end{figure}

The current LHCb detector is shown in figure~\ref{fig:lhcb}.  The beam interaction point is situated at $z=0$ on the figure, within the Vertex Locator (VELO), a silicon strip detector which reconstructs primary and secondary vertices and contributes to the tracking.  The other tracking system components are a large area silicon strip detector (TT) located in front of a 4 Tm dipole magnet, and a combination of silicon strip detectors (Inner Tracker, IT) surrounded and straw drift chambers (Outer Tracker, OT) placed  at 3 positions behind the magnet.  Two ring-imaging Cherenkov (RICH) detectors are used to identify charged hadrons.  Further downstream an Electromagnetic Calorimeter (ECAL) is used for photon detection, followed by a Hadron Calorimeter (HCAL), and a system consisting of alternating layers of iron and chambers (MWPC and triple-GEM), that distinguishes muons from hadrons (Muon System).  The ECAL, HCAL, and Muon System, together with two dedicated stations in the VELO (the Pile-Up detectors) are equipped to provide information to the current LHCb first-level trigger.

\subsection{Pileup and Occupancy}

\begin{figure}[htb]
\centering
\includegraphics[width=0.45\textwidth]{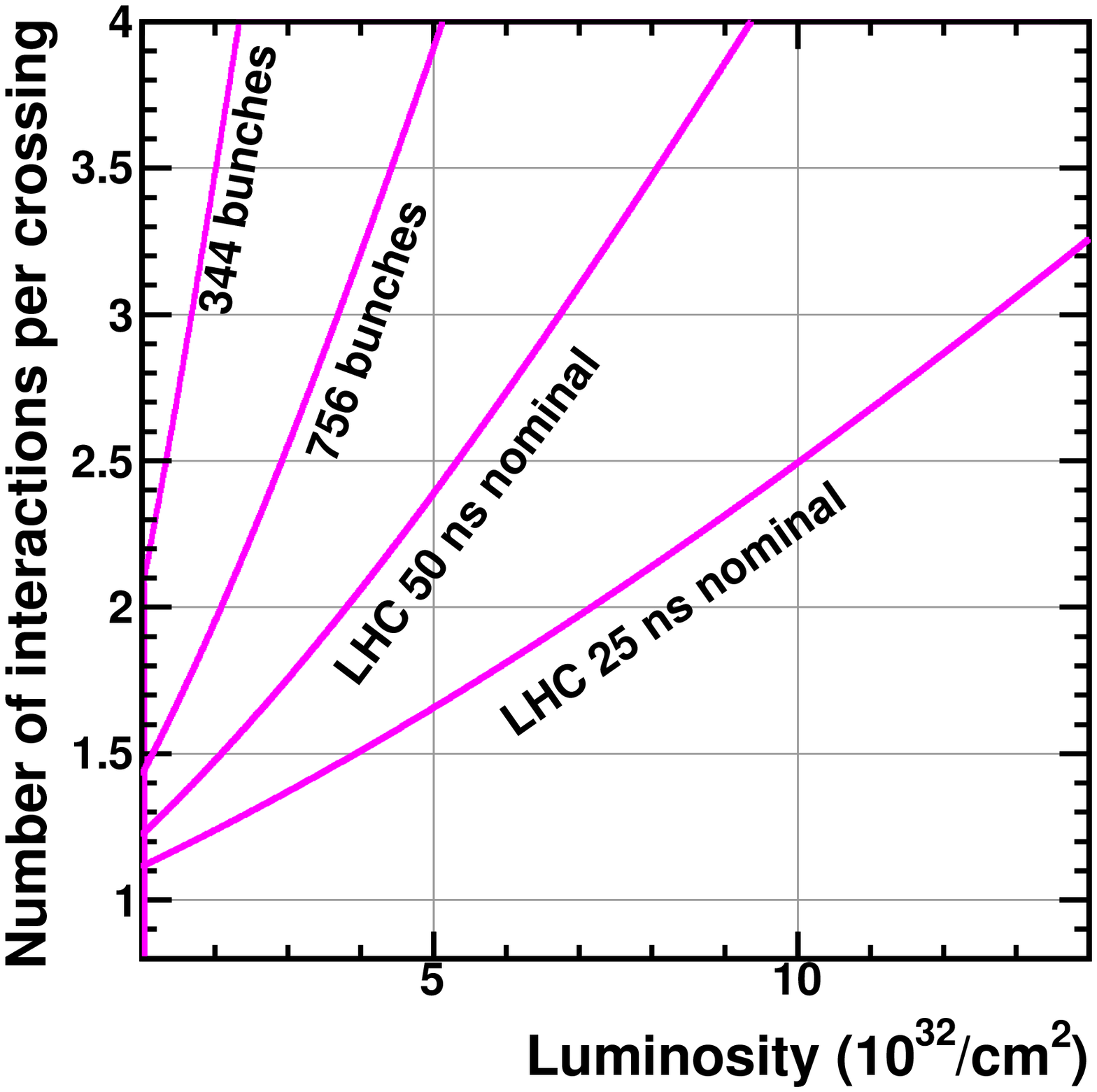}
\includegraphics[width=0.45\textwidth]{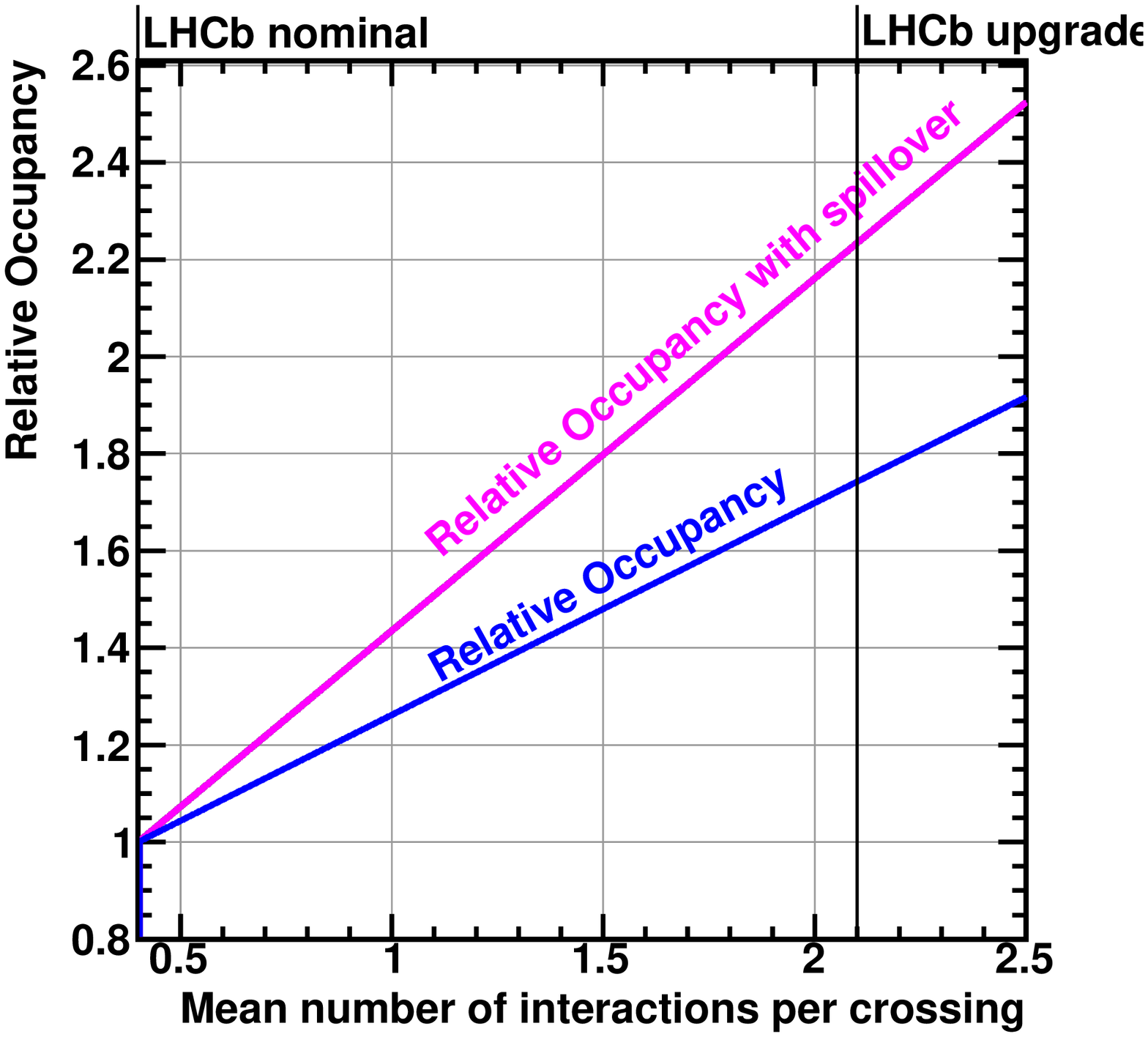}
\caption{The left plot shows the Pileup (mean number of interactions in visible events) as a function of the luminosity, for different numbers of colliding bunches at LHCb corresponding to the evolving LHC filling schemes.  In 2010 there were typically around 350 bunches, in 2011 there were up to 1380 bunches.  The right plot shows the evolution of the occupancy in $B_s \rightarrow \phi \phi$ signal events as a function of mean number of events per crossing, relative to the LHCb nominal conditions, for detectors insensitive to and sensitive to spillover.}
\label{fig:pileup}
\end{figure}

During 2010 and 2011 significant experience has been acquired in LHCb in running conditions which are similar to those expected at the upgrade.  The experiment has been running at luminosities close to or above design luminosity but with the number of bunches in the machine below nominal, i.e. with bunch spacings greater than 25~ns.  The effect on this on the pile-up, defined as the number of interactions per triggered bunch crossing is shown in figure ~\ref{fig:pileup}.  Operation has occured at pileup values of $\sim 2.5$, which is very similar to that forseen at the upgrade.  The signal to background of key channels have been investigated as a function of pileup and in general the loss of sensitivity is small.  The evolution of the reconstruction efficiency as a function of pileup and occupancy has been studied with the real data and compared to simulation, giving good confidence that the simulation can correctly estimate the detector performance at the upgrade.  The detector has not yet experienced spillover (defined as crosstalk from the previous or next event) from 25 ns running, however in 2011 the LHC has started operating with 50 ns bunch spacing, and this, together with the long drift times currently used in the outer tracker has enabled us to start to gain experience in this area also.

The evolution of the detector occupancy as a function of Pileup is shown in figure ~\ref{fig:pileup} (right) which shows the occupancy increase as a function of pileup, relative to the design pileup value of 0.4, for $B_s \rightarrow \phi \phi$ signal events.  The occupancy increase for those detectors not exposed to pileup is seen to be reasonable.  When spillover is taken into account the occupancy increase is sharper.  This is particularly relevant for the Outer tracker subdetector of LHCb, which is constructed with straw draft chambers, and eventually the drop in track reconstruction efficiency in this detector offsets the gain in luminosity, particularly for reconstruction of multibody final states.  
The LHCb experiment has already undergone a ``design reoptimisation" exercise~\cite{ref:tdrlight} to ensure that the material budget of the current detector has been kept to a minimum; at the upgrade this issue is equally important, due to the demands on keeping the occupancy as low as possible and maintaining momentum resolution, and the R\&D will focus on maintaining or improving the current material budget.

\subsection{Irradiation}

For a factor 10 increase in accumulated integrated luminosity the same factor has to be taken into account from the point of view of possible radiation damage.   The principal challenge arises at large pseudorapidity $\eta$ and affects the tracking detectors and the innermost parts of the calorimeter.  The silicon detectors will have to be replaced and the increase in irradiation and the consequences for the design and cooling will be taken into account.  In case of severe damage to the innermost ECAL modules a possibility of replacement is foreseen, in addition a series of dedicated irradiation tests with hadrons are being conducted in the LHC tunnel to evaluate the radiation resistance of the current module design at high luminosity running.  The running experience at the current detector and the observation of any aging effects with the first few $\rm{fb^{-1}}$ will be factors which guide future design choices.

\section{Detector Modifications}

As described above, the LHCb upgrade entails complete replacement of the front end electronics and the implementation of new electronics architecture.  The Muon System, Outer Tracker and calorimeters will remain with only modifications to the FE electronics, and the vessels of the RICH detectors will also remain, however the HPD photon detectors which encapsulate the current FE electronics will have to be replaced.  In addition, owing to the higher occupancies in the upgrade environment, the RICH aerogel radiator and the first muon station (M1) will be removed, and the removal of the preshower (PS) and scintillating pad detector (SPD) is also being considered.  A new component of the particle identification system based on time-of-flight (TORCH) is proposed to augment the low-momentum particle identification capabilities.  For a complete discussion of the detector modifications the reader is referred to the Letter of Intent; here the major modifications which are needed to the vertexing, PID and tracking systems are highlighted.

\subsection{VELO}

The upgraded VELO must provide a similar performance to the current VELO in terms of resolution and material budget, while providing a radiation hard design with sufficient cooling power to drain heat away from the irradiated tip of the sensors.  It is intended to reuse large most of the mechanics, vacuum systems, and evaporative $CO_2$ cooling system.   The modules and readout architecture must be completely replaced. A pixel based solution is under development, based on an evolution of the Timepix/Medipix family of chips.  The chip is based on a $55 {\rm \mu m}$ by $55 {\rm \mu m}$ square pixel, and the upgrade geometry is similar to the current VELO, with each pair of $R$ and $\phi$ strip sensors being replaced by one pixel plane.   The new front end chip, dubbed VELOPix, will send out time stamped analogue pixel information and operate in data driven mode.  One of the major challenges is to cope with the enormous data rate: a mean of 6 particles per bunch crossing for the innermost chips translates into data rates of the order of 10 Gbit/s.  The R\&D has focused on designing a chip architecture which can cope with the rate without loss of efficiency, in conjunction with a method of saving bandwidth by grouping the individual pixels into $4 \times 4$ superpixel clusters with a single timestamp.  The data will be transmitted off chip using high speed serial links via copper flex cables capable of high speed transmission to the exit of the vacuum tank.  Here electo-optical circuits will transfer the signals to optical fibres which will deliver the data to the remote backend electronics.   

\begin{figure}[htb]
\centering
\includegraphics[width=0.9\textwidth]{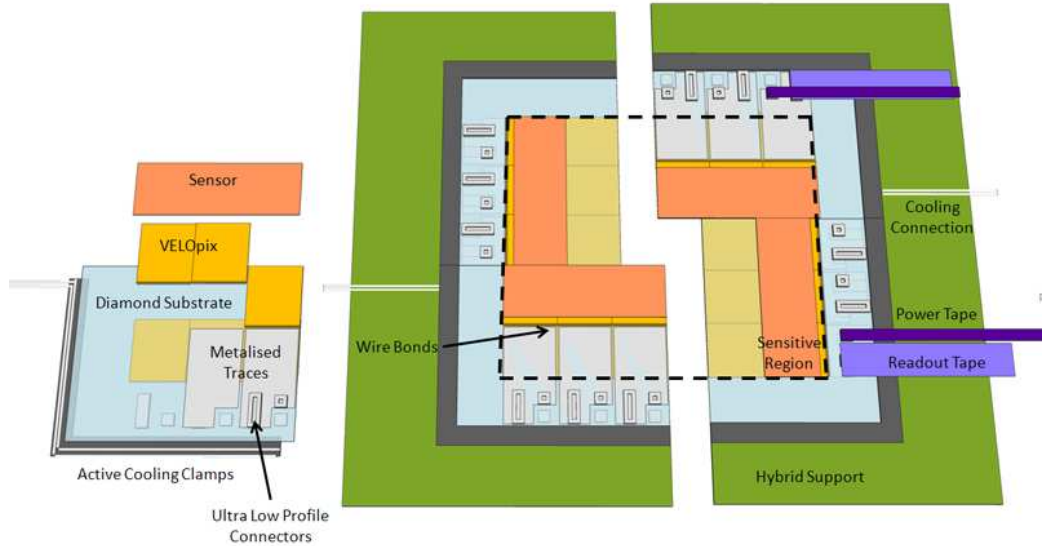}
\caption{Proposed layout of the upgraded VELO pixel module}
\label{fig:pixmod}
\end{figure}

The pixel module concept is illustrated in figure~\ref{fig:pixmod}.  Four 3-chip pixel ladders are assembled in an L-shaped arrangement on alternating sides of a diamond substrate, which acts as a cooling interface.  For the innermost chips, metal traces on the diamond bring the signals to miniaturized connectors.  The outermost part of the module, where the material is less critical, consists of a TPG (Thermal Pyrolytic Graphite) frame which acts as a support for the cooling and readout infrastructure.  Twenty-six such modules are arranged along the beam direction.  The shape of the foil, which separates the secondary vacuum of the VELO from the primary LHC vacuum must be modified to accommodate the L shaped structure.  Various options are under consideration, including use of carbon fibre composites, using workable metal alloys as for the current VELO, or direct milling of the foil from a solid metal block.   A goal of the R\&D is to achieve a reduction in material compared to the current design, which contributes significantly to the amount of material seen before the second measured point, and is an important consideration for the impact parameter resolution.  

Initial tests with the Timepix chip have demonstrated its suitability as a readout device for charged particle tracking, and a testbeam telescope has been constructed showing very promising results with the current version of the chip~\cite{ref:timepixtestbeam}.  This R\&D programme will be extended in the coming years to characterise the new versions of the ASIC and to evaluate the various sensor technologies and module prototypes being developed for the VELO upgrade.

In case the pixel solution does not meet the material and power budgets required for the VELO, a strip solution is also under development, with finer pitch and lower mass than the current VELO.  In such a case the ASIC development would be in common with any ASIC developed for a possible silicon strip readout for the upgraded TT and IT trackers.

\subsection{Tracking}

The downstream tracking in the current LHCb detector is provided by a one silicon strip tracking station (TT) in front of the magnet, and three tracking stations downstream, each consisting of large Outer Tracker (OT) straw detectors covering $98\%$ of the $30 {\rm m^2}$ detector surface, and a central cross of Inner Tracker (IT) silicon detectors, covering $0.3 {\rm m^2}$ of the surface area, but capturing $\sim 20\%$ of the tracks, at the high track density region at high $\eta$.  At the upgrade the silicon modules must be replaced, due to the incompatibility of the present electronics with the 40 MHz readout.  

The current layout is driven by occupancy considerations; the IT is designed to cover sufficient surface area such that the occupancy in the innermost parts of the straw tubes remains below $10\%$.   In practice events with higher pileup have been successfully reconstructed with the current geometry, however it is this consideration which drives the target luminosity for the first phase of LHCb upgrade.  One contributing factor to the OT occupancy is the material of the IT infrastructure inside the sensitive volume, consisting of readout hybrids and cables, mechanical supports and cooling channels.  This material has a rather non-uniform distribution and at the upgrade a reduction of material, for instance by moving the readout infrastructure outside the sensitive area, will be aimed for.  In addition, options of increasing the surface area coverage of the Inner Tracker and reducing the straw tube lengths in the regions above and below the beam pipe are under consideration, given the excellent performance of the current IT in the high pileup environment.  The radiation tolerance of the OT straws is also a consideration for the upgrade, with the current detector having been designed to accumulate an integrated luminosity of $\sim 20~{\rm fb^{-1}}$, but with some indications of aging effects emerging from laboratory tests.  The experience of 2011 running will be important to estimate the longevity of the current detector, and will give an opportunity to test the various solutions which have been proposed to overcome aging issues.

The silicon technology used by the current IT and TT tracking stations provides excellent performance and a silicon sensor based upgrade with 40 MHz electronics and suitable redesign to cope with the high luminosity operation is an attractive option for the upgrade.  For the TT, the baseline module option has been studied which integrates the electronics along the module length, reducing the input capacitance seen by the FE and allowing thinner modules and greater segmentation.  A new 130~nm CMOS ASIC must be designed with a high speed serializer capable of driving out zero suppressed data.  

\begin{figure}[htb]
\centering
\includegraphics[width=1.0\textwidth]{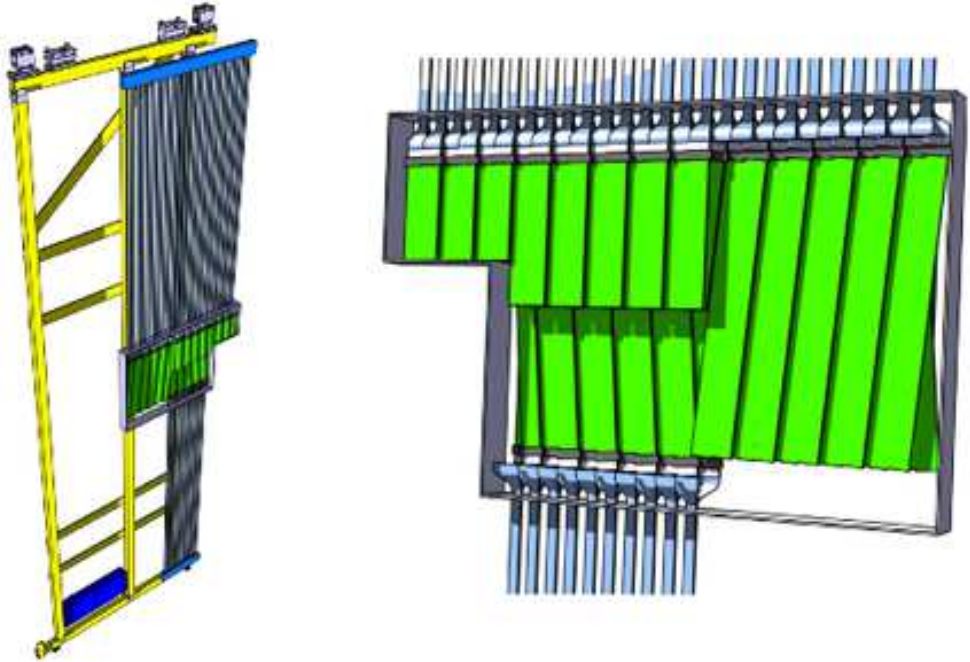}
\caption{Proposed configuration for a new IT layout based on scintillating fibres.}
\label{fig:itlayout}
\end{figure}

For a new, enlarged, Inner Tracker, the option of scintillating fibres is being pursued as an alternative to silicon.  This technology could also be deployed for the TT and, with less fine pitch, for part of the OT.  It has the advantage that the cooling requirements are less stringent than for silicon detectors, which could be particularly advantageous for high irradiation environments.  In addition the optical signals can be transported with clear fibres to readout electronics outside the detector volume.   This offers a possibility to reduce the amount of dead material inside the spectrometer acceptance.  A possible layout of a scintillating fibre IT upgrade layout is shown in figure~\ref{fig:itlayout}.  The fibres have a pitch of $250 {\rm \mu m}$ and are arranged in 5 layer stacks with a total thickness of about 1~mm.  The readout is provided with Silicon Photomultipliers with 128 sensors per chip.  The current R\&D is focussing on the mechanical assembly of such devices with sufficient precision, the use of the clear fibres to transmit the signal, and the radiation hardness of this option.

\subsection{Particle Identification (PID)}

The physics goals of the upgraded LHCb detector will rely on the continued availability of a powerful particle identification system.  The two existing RICH detectors will be re-used and most of the mechanical and optical components will remain untouched.  The Pixel HPD photon detectors, which encapsulate the electronics, are not compatible with the 40 MHz readout and must be replaced.  The baseline solution is to use multi-anode photomultipliers with external 40 MHz readout electronics; new HPDs with external readout are also under consideration, as well as the use of a novel hybrid Micro-Channel Plate photon detector (MCP) with a Medipix/Timepix readout chip similar to that used for the VELO.  Low momentum PID is currently provided by aerogel and $C_4 F_{10}$ radiators in the upstream RICH-1 detector.  For the upgrade the occupancy of photodetector hits for the aerogel component will reach unacceptable levels and it is planned to remove the radiator completely, saving $5\%~X_0$ in the detector acceptance.  The low momentum PID will be augmented with a novel time-of-flight system, the TORCH, based on the detection of Cherenkov light produced in a $1~{\rm cm}$ thick quartz plate situated downstream in the experiment.  The produced photons propagate by total internal reflection to the edge of the quartz plate and are focussed onto an array of MCPs at the periphery.  The time of arrival of the photons depends on the track time of flight together with the Cherenkov angle dependent path length in the quartz plate.  It is anticipated that with a 70~ps resolution per single-photon time measurement, sufficient separation power can be achieved for the physics goals of the experiment, a target that seems reachable with the use of MCPs in combination with the NINO and HPTDC chip sets.  

\section{Conclusions}

The LHCb experiment is operating very successfully in a high multiplicity environment at the LHC. After accumulating $\sim 5~{\rm fb^{-1}}$ over the next 5 years the experiment will be upgraded to allow running at higher luminosity and to enhance the data taking rate by an order of magnitude.  The experiment will be read-out at 40 MHz and the trigger algorithms will be executed in software.  Over 10 years an integrated luminosity of $50~{\rm fb^{-1}}$ will be accumulated with an improved trigger efficiency.  The upgrade is not dependent on any LHC luminosity upgrade and can be installed in the LHC shutdown foreseen for 2018.   The current vertexing, PID, tracking and triggering performance give confidence that the upgrade will be successful.  It will allow LHCb to reconstruct huge samples of b and c hadron decays, with unique access to the $B_s$ sector.  The LHCb upgrade is well placed to act as a next generation flavour experiment, and the upgrade also provides unique and complementary capabilities for New Physics studies beyond flavour physics.




\end{document}